\documentclass[aps,prc,twocolumn,superscriptaddress,showpacs]{revtex4-2}

\usepackage[pdftex]{graphicx}
\usepackage{amsmath} 
\usepackage{amssymb} 
\usepackage{mathtools}
\usepackage{color}
\usepackage{hyperref}
\usepackage{textcomp} 
\usepackage{lmodern} 
\usepackage[T1]{fontenc}

\begin{document}

\title{Scattering of ultracold neutrons from rough surfaces of metal foils} 

\author{Stefan D\"{o}ge}
\email[Corresponding author Stefan Doege: ]{stefan.doege@tum.de}

\affiliation{Technische Universit\"{a}t M\"{u}nchen, Department of Physics E18, James-Franck-Strasse 1, D-85748 Garching, Germany}
\affiliation{Institut Laue--Langevin, 71 avenue des Martyrs, F-38042 Grenoble Cedex 9, France}
\affiliation{Frank Laboratory of Neutron Physics, Joint Institute of Nuclear Research (JINR), 6 Joliot-Curie Street, Dubna, Moscow Region, Ru-141980, Russia}

\author{J\"{u}rgen Hingerl}
\affiliation{Technische Universit\"{a}t M\"{u}nchen, Department of Physics E18, James-Franck-Strasse 1, D-85748 Garching, Germany}
\affiliation{Institut Laue--Langevin, 71 avenue des Martyrs, F-38042 Grenoble Cedex 9, France}

\author{Egor V. Lychagin}
\affiliation{Frank Laboratory of Neutron Physics, Joint Institute of Nuclear Research (JINR), 6 Joliot-Curie Street, Dubna, Moscow Region, Ru-141980, Russia}
\affiliation{Lomonosov Moscow State University, GSP-1, Leninskie Gory, Moscow, Ru-119991, Russia}
\affiliation{Dubna State University, Universitetskaya Street 19, Dubna, Moscow region, Ru-141982, Russia}

\author{Christoph Morkel}
\affiliation{Technische Universit\"{a}t M\"{u}nchen, Department of Physics E21, James-Franck-Strasse 1, D-85748 Garching, Germany}

\begin{abstract}
The transparency of metal foils for ultracold neutrons (UCNs) plays an important role in the design of future high-density UCN sources, which will feed a number of fundamental physics experiments. In this work, we describe and discuss the measured transmission of a collimated beam of very slow neutrons (UCNs and very cold neutrons) through foils of Al, Cu, and Zr of various thicknesses at room temperature. Our goal was to separate scattering and absorption in the sample bulk from surface scattering, and to quantify the contribution of the surface. We were able to demonstrate that the surface roughness of these foils caused a significant fraction of UCN scattering. The surface roughness parameter $b$ extracted from UCN measurements was shown to be of the same order of magnitude as the surface parameter determined by atomic-force microscopy. They lie in the order of several hundreds of angstroms. Using the formalism developed here, transmission data from previous neutron-optical experiments were re-analyzed and their surface roughness parameter $b$ was extracted.
\\
\\
Published online on 9 December 2020: \url{https://doi.org/10.1103/PhysRevC.102.064607} \\ S. D\"{o}ge et al., J. Hingerl, E. V. Lychagin, C. Morkel, Physical Review C 102 (6), 064607 (2020) \\ \textcopyright\, 2020. This manuscript version is made available under the \href{https://creativecommons.org/licenses/by/4.0/}{CC-BY 4.0 license}.
\end{abstract}

\pacs{03.75.Be, 25.40.Dn, 28.20.Cz, 83.85.Hf} 

\maketitle

\section{Introduction}

Metal foils are frequently used in experiments where ultracold neutrons (UCNs) need to pass but the vacuum of different volumes must be separate, e.g., the vacuum at the PF2 instrument ``Turbine'' at the Institut Laue--Langevin in Grenoble, France,~\cite{steyerl:1986,doege:2020-turbine}, and the neutron beam guide vacuum. They also play an important role as neutron exit windows in the exploitation of other UCN sources~\cite{ries:2017,ito:2018,ahmed:2019}, such as those based on solid deuterium or liquid helium. To extract a maximum UCN flux from these sources, UCN losses on these exit windows need to be as low as possible.

The fact that the surface roughness of foils has a tremendous effect on UCN transmission through media has been shown by Steyerl~\cite{steyerl:1972} and Roth~\cite{roth:1977} but has, apparently, found no attention so far.

Studies of exit window materials for the UCN sources ``Mini-D2''~\cite{frei:2002} in Mainz, Germany, and the one at the Paul Scherrer Institut~\cite{atchison:2009-foils}, Switzerland, investigated the best choice of materials for this application. Aluminum and zirconium were identified as the best candidates due to their low neutron absorption cross section. These studies used thin metal foils in the range of a few hundred micrometers thickness. They noticed a significantly lower transmission of UCNs than expected. Besides neutron absorption by strongly absorbing isotopes present as trace impurities in the samples, surface scattering was conjectured to be one of the causes but no quantitative explanation or estimate was provided.

As Lavelle et al.~\cite{lavelle:2010} demonstrated experimentally, the rough aluminum windows of their sample container caused UCN losses of up to 3/4. Alas, this effect was not further studied and has not been taken into account quantitatively in the literature on UCN \emph{transmission} experiments, in which rough-surface sample containers were used. Having recognized this, we developed and used a low-roughness sample container~\cite{doege:2018}. The present paper gives an impression of the magnitude of measurement uncertainty for UCN cross sections when rough sample containers are used instead of low-roughness containers.

The following sections deal with the purpose of this paper (Sec.~\ref{sect:idea-of-experiment}), UCN losses in the sample bulk and their calculation (Sec.~\ref{sect:losses-bulk}), and UCN losses on the sample surface (Sec.~\ref{sect:ucn-loss-surface}), which are calculated as the difference between the measured total UCN losses and the known loss channels in the sample bulk. The experimental setups for both the neutron and AFM measurements are given (Sec.~\ref{sect:exp-setup}), followed by a discussion of the results, their application to previously performed experiments, and their implication for UCN transmission measurements in general (Sec.~\ref{sect:results-discussion}).

\section{Idea of the Experiment}\label{sect:idea-of-experiment}

The laws of neutron optics govern the transmission of ultracold neutrons through metal foils. We hypothesize that the foil thickness has only a very limited influence on the total transmission -- and thus the loss -- of UCNs. Much more important is the surface roughness of the foil.

The exact interaction of rough surfaces with neutron beams depends on the beam distribution -- collimated, isotropic, or some other shape. For exit foils and UCN transmission experiments, the attenuation of collimated beams is important and thus the subject of the present paper. In this context, a UCN is considered lost when it is absorbed by a nucleus of the sample bulk, gains energy through an \emph{up-scattering} event and leaves the UCN energy range, or when it is diverted from the collimated neutron beam by elastic scattering. In UCN storage experiments, for example, neutrons impinging on the surface of storage vessels are not collimated and undergo both specular and diffuse reflection. Their loss is only due to absorption or up-scattering by the vessel's walls~\cite{ignatovich:1990,pokotilovski:1999} or by impurities on the wall's surface~\cite{strelkov:1978,lychagin:2000}. However, surface roughness can, depending on the conditions, decrease or increase the probability of UCN loss by absorption upon a collision with the wall~\cite{steyerl:2010}.

In Frei~\cite{frei:2002} and Atchison et al.~\cite{atchison:2009-foils}, the different sample thicknesses were achieved by layering several thinner foils on top of each other. By doing that, the thickness of the bulk material was increased, but so was the number of surfaces. It was, therefore, not surprising that the thickness-dependent transmission curves showed an exponential decay with increasing foil thickness, or better: with an increasing number of foils.

To gain a better understanding of UCN scattering on rough surfaces, UCN transmission experiments through metals foils of the same bulk material and surface preparation but with different thicknesses needed to be carried out. This way, the bulk thickness -- with UCN loss cross sections known from the literature -- could be varied while the number of surfaces remained constant.

\section{Ultracold Neutron Losses in the Sample Bulk}\label{sect:losses-bulk}

\subsection{Sample impurities and oxidation}

We chose high-purity metal foils as samples for our experiments to avoid elastic scattering of UCNs on bulk inhomogeneities. The impurities of the various metal foils (Cu, Al, Zr) investigated in this work were taken from the supplier's (Advent Research, UK) list of typical impurities, which gave a good estimate of the trace impurities to be expected.

Taking into account the abundance of these impurities as well as their respective absorption cross sections, only one or two impurities per sample actually influenced the total absorption cross section. Table~\ref{tab:foil-add-absorption} lists for each sample its respective most relevant impurities. Since the impurities were very dilute, it was assumed that they had the same particle density as their host material multiplied with their respective number concentration. The absorption cross sections at thermal-neutron energies were taken from Sears~\cite{sears:1992}. The resulting sum of bulk and impurity absorption cross sections was also calculated for thermal-neutron energies. For all samples, the additional absorption cross sections due to impurities were less than 4\% and they were, therefore, neglected.

\begin{table}[!ht]
\centering
  \begin{tabular}{c c c c c}\hline\hline
Sample & Element & Number & $\sigma_\text{abs}^\text{thermal}$ & $\sigma_\text{abs}^\text{thermal}$ \\
type & & conc. (\%) & (b) & (b) weighted \\\hline

Al 99.99+\%& Al & > 99.99 & 0.231 & 0.231 \\
 & Ag & $3\times 10^{-6}$ & 63.3 & $1.9\times 10^{-4}$ \\
 & B & $2\times 10^{-7}$ & 767 & $1.5\times 10^{-4}$ \\
\textit{Sum} & & & & \textit{0.231} \\\hline

Cu 99.9+\%& Cu & > 99.9 & 3.78 & 3.78 \\
 & Ag & $5\times 10^{-4}$ & 63.3 & $3.2\times 10^{-2}$ \\
\textit{Sum} & & & & \textit{3.81} \\\hline

Zr 99.8\%& Zr & 99.8 & 0.185 & 0.185 \\
 & Fe & $8.4\times 10^{-4}$ & 2.56 & $2.2\times 10^{-3}$ \\
 & Hf & $4.4\times 10^{-5}$ & 104 & $4.6\times 10^{-3}$ \\
\textit{Sum} & & & & \textit{0.192} \\\hline\hline

  \end{tabular}
\caption[Relevant impurities in metal foils]{Sample impurities giving rise to additional absorption cross sections in the metal foil samples i) Al 99.99+\%, ii) Cu 99.9+\%, and iii) Zr 99.8\%. The ``+'' denotes a sample purity even higher than indicated. All cross sections are given for thermal-neutron energies.}\label{tab:foil-add-absorption}
\end{table}

The foils used here had received the following surface treatments: Al -- temper as rolled, Cu and Zr -- temper annealed.

When metals like aluminum and copper are exposed to air at ambient temperature, they form passive oxide layers. For pure bulk copper, a layered CuO/CuO$_2$ oxide structure of 3.3~nm thickness has been reported~\cite{keil:2007}, while 2.5 to 5.2~nm were found for copper thin films~\cite{platzman:2008} and 6~nm for ultrafine particles~\cite{tamura:2003}. The oxide layer on the surface of pure aluminum has been found to be 4~nm for ultrafine particles~\cite{tamura:2003} and between 3 and 4~nm for bulk aluminum~\cite{evertsson:2015}. On zirconium, ZrO$_2$ and substoichiometric oxides have a thickness of about 1.5~nm~\cite{bespalov:2015}.

These oxide layers are thinner than one typical UCN wavelength, which is a few hundred angstroms. According to the UCN reflectivity calculations by Poko\-ti\-lov\-ski~\cite{pokotilovski:2016}, they are thin enough to have only negligible influence on the UCN transmission through the sample.

\subsection{One-phonon up-scattering}

The total UCN cross section of a sample is composed of bulk scattering, surface scattering, and absorption in the bulk. From Table~\ref{tab:foil-add-absorption}, the absorption cross sections at thermal-neutron energies ($E_\text{kin}=25$ meV, $v=2200$ m/s) can be taken and extrapolated to UCN energies by using the relation $\sigma_\text{abs}\times v = \text{const.}$, see for example Ignatovich~\cite{ignatovich:1990}.

Table~\ref{tab:foil-scattering-absorption} lists the one-phonon up-scattering cross sections $\sigma_\text{1-ph}^\text{IA}$ calculated for room temperature using the Incoherent Approximation (IA)~\cite{turchin:1965}, the one-phonon up-scattering cross sections $\sigma_\text{1-ph}^\text{corrIA}$ taking into account corrections to the Incoherent Approximation (corrIA) by Placzek and Van Hove~\cite{placzek:1955}, the absorption cross sections $\sigma_\text{abs}$ \cite{sears:1992}, and the total UCN loss cross section $\sigma_\text{tot}$ for all three metals under investigation -- Al, Cu, and Zr. The corrections to the IA for coherent scatterers can only strictly be calculated for cubic crystals. Since Zr has a hexagonal close-packed (hcp) crystal structure at room temperature, it was assumed to have a face-centered cubic (fcc) structure for the sake of calculating the correction. For all three metals at room temperature, one-phonon up-scattering is not significant compared to absorption, which is the dominant UCN loss channel. All cross sections listed in Table~\ref{tab:foil-scattering-absorption} can be extrapolated to other neutron energies using the relation $\sigma \times v = \text{const.}$

\begin{table}[!ht]
\centering
  \begin{tabular}{c c c c c}\hline\hline
Sample & $\sigma_\text{1-ph}^\text{IA}$ (b) & $\sigma_\text{1-ph}^\text{corrIA}$ (b) & $\sigma_\text{abs}^\text{8m/s}$ (b) & $\sigma_\text{tot}^\text{8m/s}$ (b)\\\hline
Al & 7.9 & 8.8 & 63.5 & 72.3\\
Cu & 23.5 & 21.6 & 1040 & 1062\\
Zr & 15.9 & 16.7 & 50.9 & 67.6\\\hline\hline
  \end{tabular}
\caption[UCN up-scattering vs. absorption in metal foils]{One-phonon up-scattering calculated according to the Incoherent Approximation (IA), IA with corrections for coherent effects (corrIA), as well as absorption for pure Al, Cu, and Zr, taken from Table~\ref{tab:foil-add-absorption}. The right-hand column gives the total UCN loss cross section as the sum of $\sigma_\text{1-ph}^\text{corrIA}$ and $\sigma_\text{abs}$. All values were calculated for room temperature and an in-medium velocity of $v_\text{inm} = 8\,\text{m/s}$ of the UCNs.}\label{tab:foil-scattering-absorption}
\end{table}

Coherent elastic scattering in the bulk can be neglected due to the UCNs' large wavelengths, which go far beyond any Bragg cutoff wavelength, and due to the high sample purity, which avoids scattering on inhomogeneities in the sample bulk. With up-scattering and absorption in the sample bulk being known quantities, the remaining loss of UCNs in the transmission experiment on metal foils can be attributed to surface scattering only.

\section{Ultracold Neutron Loss on Surfaces}\label{sect:ucn-loss-surface}

\subsection{Separating bulk from surface losses}\label{sect:separate-ucn-loss-surf}

The standard transmission equation for uniformly absorbing and scattering media (in optics known as the Lambert--Beer law~\cite{bouguer:1729, beer:1852}),
\begin{equation}
T = \frac{I_\text{n}}{I_0} = \text{e}^{-N\sigma_\text{tot} d_\text{n}},
\end{equation}

\noindent
can be expanded by $1-L_\text{t}$ to account for surface scattering~\cite{steyerl:1972},
\begin{equation}\label{eq:transmission-formula-mod}
T = \frac{I_\text{n}}{I_0} = \underbrace{\text{e}^{-N\sigma_\text{tot} d_\text{n}}}_{(1-A)} \times (1-L_\text{t}),
\end{equation}

\noindent
where $T$ represents the measured absolute transmissivity of sample $n$, $I_0$ the incoming neutron beam intensity, $I_\text{n}$ the neutron beam intensity behind sample $n$, and $\sigma_\text{tot}$ (for in-medium neutron velocity) the total UCN loss cross section of the sample bulk, as discussed in Sec.~\ref{sect:losses-bulk}. $A$ is the loss of UCNs in the sample bulk (between the two surfaces) and $L_\text{t}$ is the integral probability of diffuse scattering from rough surfaces, i.e., the UCN loss on the two surfaces of one foil. The particle number density $N$ is known from the literature and the sample thickness $d_\text{n}$ can easily be measured. To determine $L_\text{t}$, the transmissivity $T$ of a foil needs to be measured. Then, Eq.~\ref{eq:transmission-formula-mod} can be solved for $L_\text{t}$,
\begin{equation}\label{eq:transmissivity-two}
L_\text{t} = 1- \frac{T}{\text{e}^{-N\sigma_\text{tot} d_\text{n}}}.
\end{equation}

\subsection{Connecting surface roughness and ultracold-neutron Loss}\label{sect:surface-rough-ucn-loss}

For UCNs that are incident perpendicularly on a rough sample surface (``macroroughness'') and have a wave vector (out of medium) $k\gg \left|k_\text{l}\right|$, Steyerl~\cite{steyerl:1972} defined the total fraction of UCNs that is scattered out of the direct beam by the two surfaces of a sample as
\begin{equation}\label{eq:steyerl-surface}
L_\text{t} = \frac{I_\text{2-surf}}{I_0} = \frac{1}{4}\frac{b^2 k_\text{l}^4}{k^2} \left(1 + \text{e}^{-N\sigma_\text{tot} d_\text{n}}\right),
\end{equation}
\noindent
where $k_\text{l}=m_\text{n} \times v_\text{crit}/\hbar $, $m_\text{n}$ is the neutron's mass, $v_\text{crit}$ is the critical velocity of the sample, $b$ is the surface roughness parameter, $I_0$ is the neutron beam intensity incident on the rough surface, and $I_\text{2-surf}$ is the intensity of UCNs scattered out of the direct beam by the two surfaces. The term $1 + \text{e}^{-N\sigma_\text{tot} d_\text{n}}$ (instead of 2) accounts for the loss of UCNs in the sample bulk between the two scattering surfaces. The first surface receives the full incoming neutron intensity, the surface downstream from it sees a beam attenuated by the losses in the bulk -- and on the first surface. The condition $k\gg \left|k_\text{l}\right|$ is fulfilled for the metals treated here for all except the very slow UCNs of $v \lesssim 5$~m/s (out of medium).

The surface roughness parameter $b$ as seen by the UCNs can thus be derived by solving Eq.~\ref{eq:steyerl-surface}
\begin{equation}\label{eq:steyerl-b}
b = \sqrt{\frac{4 L_\text{t} k^2}{k_\text{l}^4 \left(1 + \text{e}^{-N\sigma_\text{tot} d_\text{n}}\right)}}.
\end{equation}

\noindent
This parameter $b$ is defined as the mean square amplitude of elevations above and below the reference plane of the surface. For surfaces with a relatively even distribution of peaks and valleys and no extreme peaks, the mean square amplitude is quite similar to the center-line average roughness $R_\text{a}$, which we measured using atomic-force microscopy, see Sec.~\ref{sect:afm}.

\section{Experiment Setup}\label{sect:exp-setup}

\subsection{Ultracold-neutron transmission experiments}

For the transmission measurements with very slow neutrons with an out-of-medium velocity $v_\text{oom}$, $3 \leq v_\text{oom} \leq 15$ m/s, we used the time-of-flight (TOF) method and a collimated beam of UCNs and very cold neutrons (VCN) at the PF2-EDM beamline~\cite{doege:2020-turbine} of the Institut Laue--Langevin. The neutron beam was strongly collimated in the forward direction with a solid angle of the collimator aperture of $\Omega = 2.2 \times 10^{-2}$~sr. The TOF geometry used in these experiments is explained in detail in D\"{o}ge et al.~\cite{doege:2015}. As sample holder for the metal foils we used the one that was developed for UCN transmission experiments on liquid and solid deuterium~\cite{doege:2018}. Aluminum clamps held the metal foils in place during the transmission experiments.

The foil samples were cleaned with high-purity ethanol and dried immediately prior to their installation into the vacuum vessel where the measurements took place. Before the UCN measurements were started, the vacuum was stabilized in the 10$^{-3}$~mbar range for half an hour. The measurements themselves ran over several hours and showed no significant fluctuation in the UCN transmission over time as the vacuum continuously improved to the lower 10$^{-4}$~mbar range. It can therefore be considered certain that all water and volatile compounds, which may have been adsorbed onto the metal surface between the installation of the sample foil in the vacuum chamber and the start of the evacuation, evaporated and caused no additional scattering of UCNs.

\subsection{Atomic-force microscopy measurements of surface roughness}\label{sect:afm}

The mechanical surface roughness of the metal foils was measured at JINR Dubna under an atomic-force microscope (AFM) from NT-MDT. The center-line average roughness, i.e., the average deviation from the imaginary center plane of the surface~\cite{whitehouse:2004},

\begin{equation}
R_\text{a} = \frac{1}{n}\sum_{i=1}^{n} \left| y_i \right|
\end{equation}

\noindent
for each foil sample was determined from two or more two-dimensional scans of 5~$\mu$m $\times$ 5~$\mu$m, which were carried out on flat sections of overview scans of 30~$\mu$m $\times$ 30~$\mu$m. This way, we minimized the role that long-wavelength surface waviness plays in the calculation of the surface roughness according to the standards ISO 4287-1:1984 and GOST 25142-82. The AFM was calibrated by using two different calibration samples made of SiO$_2$ with step sizes of 21.5~nm (TGZ1) and 107~nm (TGZ2). These samples were also used to verify the reliability of the two-dimensional roughness calculation algorithm. The global uncertainty of the measured surface roughness parameters was $\pm 30$\%. The results are given in Table~\ref{tab:foil-transmissivity}. The roughness values of all three copper foils are very close to one another. For aluminum and zirconium the mutual agreement of the roughness scans is less pronounced but still generally established. All of the roughness amplitudes are in the range of a few tens of nanometers, which is a typical wavelength of ultracold neutrons.

\begin{table}[!ht]
\centering
  \begin{tabular}{c c c c}\hline\hline
Sample & Foil Thickness & Transmissivity & Roughness $R_\text{a}$\\
 & ($\mu$m) & at $v_\text{oom}=7$ m/s & (nm)\\\hline
Al-1 & 50 & 0.967 $\pm$ 0.058 & 25.5\\
Al-2 & 100 & 0.859 $\pm$ 0.050 & 34.0\\
Al-3 & 125 & 0.851 $\pm$ 0.046 & 48.8\\\hline
Cu-1a & 50 & 0.260 $\pm$ 0.023 & 60.5\\
Cu-1b & 2$\times$50 & 0.081 $\pm$ 0.017 & 60.5\\
Cu-2 & 100 & 0.109 $\pm$ 0.014 & 61.1\\
Cu-3 & 250 & 0.010 $\pm$ 0.004 & 72.7\\\hline
Zr-1 & 25 & 0.613 $\pm$ 0.040 & 33.4\\
Zr-2 & 125 & 0.277 $\pm$ 0.024 & 37.5\\
Zr-3 & 250 & 0.310 $\pm$ 0.023 & 48.2\\\hline\hline
  \end{tabular}
\caption[Transmissivity for all metal foil samples]{The transmissivity for all metal foil samples is given as fraction of the direct beam as recorded during the UCN transmission experiments at the Institut Laue--Langevin, Grenoble. The surface roughness of metal foils is given as center-line average as measured by AFM. The global uncertainty for all $R_\text{a}$ values is $\pm 30$\%.}\label{tab:foil-transmissivity}
\end{table}

\section{Results and Discussion}\label{sect:results-discussion}

\subsection{Ultracold-neutron transmission measurements}

\begin{figure}[!ht]
\centering
\includegraphics[width=1.00\columnwidth]{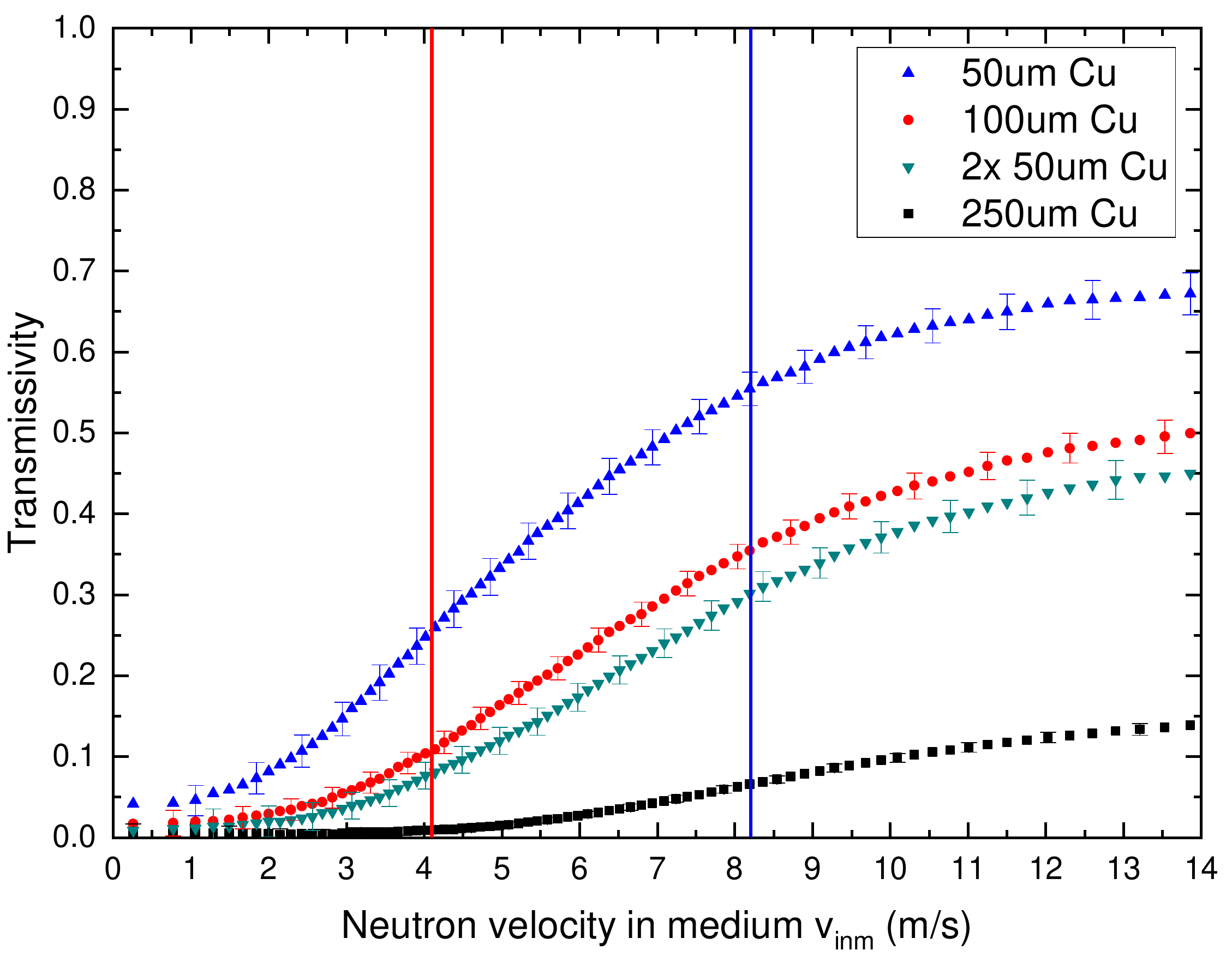}
\caption[Transmissivity of Cu foil for UCNs over UCN velocity]{\label{fig:transmissivity-cu}Measured transmissivity of Cu foils for UCNs plotted over UCN velocity. The in-medium velocity of 4.1 m/s (equivalent to 7 m/s out of medium) is marked with a vertical red line. The vertical blue line marks 8.2 m/s (equivalent to 10 m/s out of medium). For data treatment, a sliding average of 16 time bins was used. The error bars account for this. To improve legibility, only every fourth error bar is shown.}
\end{figure}

The transmission of UCNs and VCNs through metal foils of different thicknesses but with the same surface treatment was carefully recorded~\cite{doege:2018-3-14-380}, see Figure~\ref{fig:transmissivity-cu} for the copper samples. It is worth noting that the UCN transmission through a stack of two 50-$\mu$m-thick copper foils is lower than that through a single foil of 100~$\mu$m thickness. This is a direct indication of UCN losses on the two additional surfaces. Table~\ref{tab:foil-transmissivity} gives the UCN transmissivity of each metal foil for an out-of-medium (oom) UCN velocity $v_\text{oom}$ of 7 m/s that was corrected for the critical velocity $v_\text{crit}$ of each sample, see for example Ignatovich~\cite{ignatovich:1990},

\begin{equation}
v_\text{inm} = \sqrt{v_\text{oom}^2 - v_\text{crit}^2}.
\end{equation}

\noindent
This logic was chosen in keeping with Atchison et al.~\cite{atchison:2009-foils} to make their results comparable with those presented here. The resulting in-medium (inm) velocities $v_\text{inm}$ were 6.2 m/s for Al, 4.1 m/s for Cu, and 5.8 m/s for Zr.

\begin{figure}[!ht]
\centering
\includegraphics[width=1.00\columnwidth]{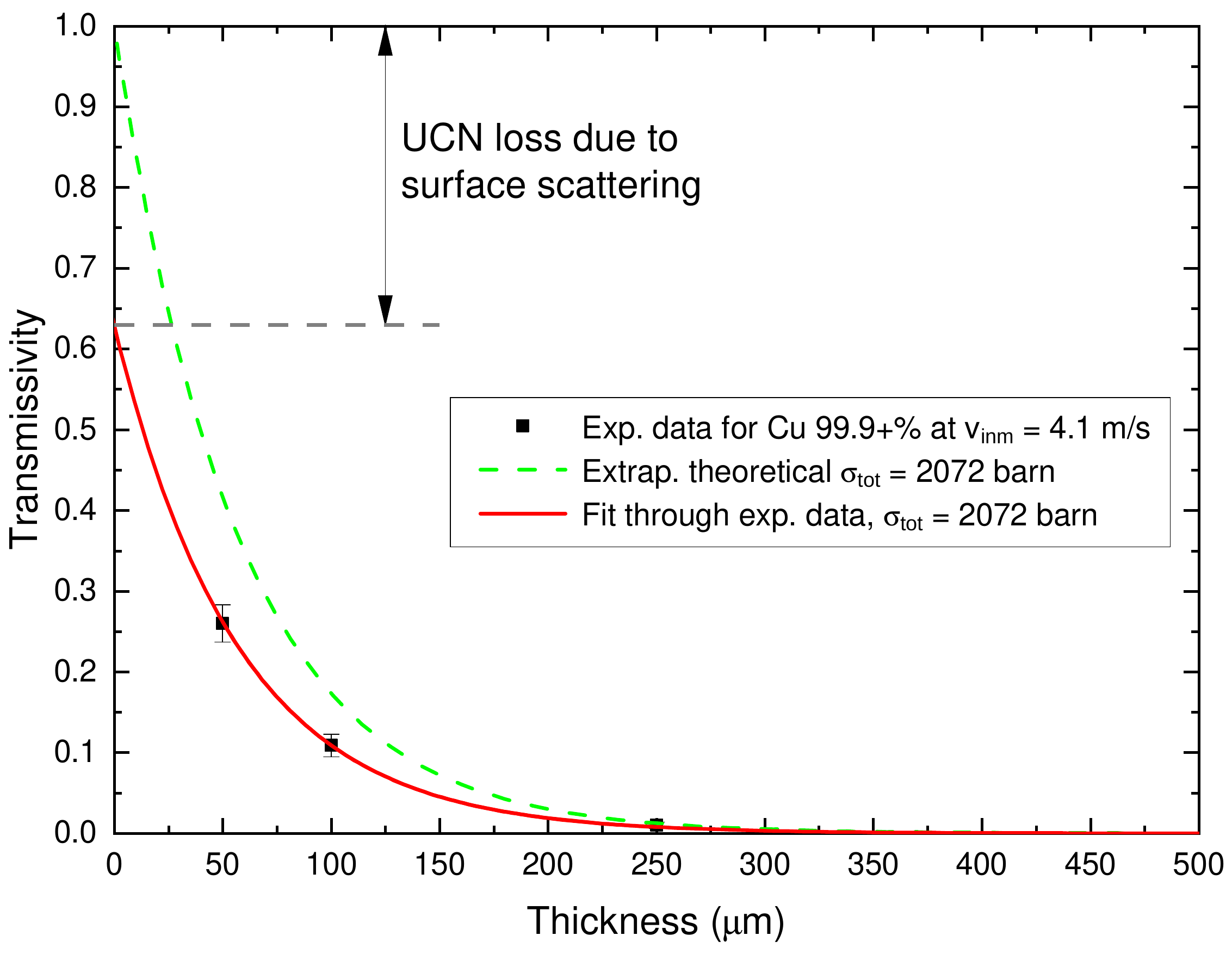}
\caption[Transmissivity of Cu foils for UCNs over thickness]{\label{fig:transmissivity-cu-thickness}The measured transmissivity of Cu foils of three different thicknesses is plotted for UCNs of an in-medium velocity of 4.1 m/s (equivalent to 7 m/s out of medium). The red line represents a fit to the data using the transmission equation including a surface scattering term, see Eq.~\ref{eq:transmission-formula-mod}. The dashed green line represents the transmissivity as expected if the sample caused no surface scattering. For both curves, $\sigma_\text{tot} = 2072$ b (at 4.1~m/s in medium) was used as UCN loss cross section of the sample bulk.}
\end{figure}

Figure~\ref{fig:transmissivity-cu-thickness} shows the UCN transmissivity of copper foils of different thicknesses. When the data points at $v_\text{inm}=4.1$~m/s are fit using the modified transmission equation, Eq.~\ref{eq:transmission-formula-mod}, and an extrapolation to zero foil thickness is done, the intercept of the fit with the $y$ axis shows the fraction of UCNs that is lost on both surfaces together, $L_\text{t}$. It is obvious that a very large share of the total UCN losses is due to surface scattering instead of bulk scattering and absorption. In the case of the copper foils, the two surfaces alone scatter about 37\% of UCNs out of the direct beam.

When the transmissivity values are taken from Table~\ref{tab:foil-transmissivity} and plugged into Eq.~\ref{eq:steyerl-b}, one can calculate the foils' roughnesses as seen by the UCNs. These roughness values are shown as parameter $b$ in Table~\ref{tab:foil-roughness-comp}. As the Al-1 sample was very transparent and the Cu-3 sample very opaque, these extremes were omitted when calculating the average sample surface roughness.

As a test, the roughnesses were also calculated using the transmissivity values at a neutron velocity of $v_\text{oom} = 10$~m/s for all three metals. They deviated only between $-18$\% and $+2$\% from the values for 7~m/s, which shows that the foil roughness as seen by neutrons is consistent over the UCN velocity range. The copper samples Cu-1a and Cu-1b, which consisted of the same type of 50~$\mu$m foils, yielded roughness values within a few percent of each other, both for 7~m/s and 10~m/s neutrons. This is another confirmation of the reliability of our approach.

\subsection{Surface roughness}

The surface roughness of the metal foil samples both as measured by atomic force microscopy (AFM) and extracted from the cross-section measurements above are shown and contrasted in Table~\ref{tab:foil-roughness-comp}.

\begin{table}[!ht]
\centering
  \begin{tabular}{c c c}\hline\hline
Sample & Roughness & Roughness\\
 & Parameter $b$ (\AA) & Parameter $R_\text{a}$ (\AA)\\\hline
Al & $182\pm 8 $ & $361\pm 164$\\
Cu & $147\pm 12 $ & $648\pm 227$\\
Zr & $313\pm 64 $ & $397\pm 155$\\\hline\hline
  \end{tabular}
\caption[Surface roughness of metal foils]{Surface roughness of metal foils as extracted from UCN measurements and the theory explained in Section~\ref{sect:ucn-loss-surface} (parameter $b$), as well as the roughness measured mechanically by AFM (parameter $R_\text{a}$). The errors for the UCN measurements include statistical and systematic errors.}\label{tab:foil-roughness-comp}
\end{table}

Table~\ref{tab:foil-roughness-comp} demonstrates that the magnitudes of both roughness parameters $b$ and $R_\text{a}$ are of the same order of magnitude.

AFM scans of the foils used in the experiments described above yielded values of the same order of magnitude as reported in the work of Steyerl~\cite{steyerl:1972}, 200 to 500~\AA. In his work, Steyerl noted that the roughness values extracted from electron micrographs were in quantitative agreement with the roughness parameters extracted from neutron measurements but that it was difficult to interpret these micrographs.

Generally, it is difficult to obtain the same roughness values for measurements of the same sample that were done using different techniques. For example, the nonzero curvature radius of the stylus used in an atomic-force microscope (AFM) leads to smaller roughness readings than those obtained by noncontact optical methods~\cite{poon:1995}. One single roughness parameter is often not enough to describe the entire surface roughness with short-range and long-range correlations. An overview of roughness parameters was published by Gadelmawla et al.~\cite{gadelmawla:2002}.

Considering the above, it has to be concluded that, with the theories and techniques currently available, mechanical measurements of sample roughness can at best serve to estimate the order of magnitude of the loss of UCNs due to scattering on rough sample surfaces. The transmission of each foil, used as vacuum barrier or for a different purpose, still has to be measured with UCNs to know its exact transmissivity.

\subsection{Reexamination of previous experiments}

Using the equations explained in Sec.~\ref{sect:ucn-loss-surface}, the experimental results from Atchison et al.~\cite{atchison:2009-foils} can be reexamined to determine the surface roughness parameter $b$ for those foils as seen by UCNs. Atchison et al. were not able to confirm the exact make-up of their foil samples~\cite{daum:2018} -- stacked or single foil. For aluminum, the first sample had a thickness of 10~$\mu$m and we suspected that the others were very likely stacks of 100-$\mu$m-thick foils. The zirconium samples were 100, 250, and 500~$\mu$m thick, likely layered from 50-$\mu$m-thick foils. For multiple layers of foil, the right side of Eq.~\ref{eq:transmission-formula-mod} needs to be raised to the power of $n$, which represents the number of foils in the neutron beam. Otherwise the lower transmissivity would be erroneously attributed to $\sigma_\text{tot}$ and suggested a false higher bulk cross section. Consequently, before calculating the surface roughness parameter $b$ of \textit{one} foil, the $n$-th root needs to be taken of the surface transmissivity term $(1-L_\text{t})^n$.

\begin{table}[!ht]
\centering
  \begin{tabular}{l c c c c}\hline\hline
Sample Thick- & Exp. & Theor. & Exp. / Th. & Roughness\\
ness $d$ ($\mu$m) & Transm. & $\exp^{-N\sigma_\text{tot}d}$ & $[1-L_\text{t}]^n$ & $b$ (\AA)\\\hline
Al $1\times 10$ & 0.943 & 0.994 & 0.949 & 136\\
Al $1\times 100$ & 0.837 & 0.941 & 0.890 & 203\\
Al $2\times 100$ & 0.707 & 0.886 & 0.799 & 202\\
Al $3\times 100$ & 0.612 & 0.833 & 0.735 & 196\\
Al $4\times 100$ & 0.540 & 0.784 & 0.689 & 190\\
Al $5\times 100$ & 0.484 & 0.738 & 0.656 & 183\\\hline
Zr $2\times 50$ & 0.860 & 0.955 & 0.901 & 93.1\\
Zr $5\times 50$ & 0.680 & 0.891 & 0.764 & 96.1\\
Zr $10\times 50$ & 0.449 & 0.793 & 0.566 & 101\\\hline\hline
  \end{tabular}
\caption[Surface roughness of metal foils, PSI]{Surface roughness $b$ of individual metal foils from a foil stack as used by Atchison et al.~\cite{atchison:2009-foils} and re-analyzed applying the theory explained above. Column 4 gives the ratio of experimental transmissivity to theoretical transmissivity (due to absorption only), i.e., column 2 divided by column 3. The uncertainty of $b$ was estimated to be $\pm 15$\% due to the uncertainty of the original neutron transmission measurements.}\label{tab:foil-roughness-psi}
\end{table}

Table~\ref{tab:foil-roughness-psi} gives the experimental transmissivities of Al and Zr foils (second column) as well as the roughness parameter $b$ extracted from those measurements by solving Eqs.~\ref{eq:transmission-formula-mod} and \ref{eq:steyerl-b}. In the calculation of $\text{e}^{-N\sigma_\text{tot}d}$, $\sigma_\text{tot}$ was taken as presented in the paper~\cite{atchison:2009-foils}; one-phonon scattering was neglected. The ratio of experimental transmissivity to theoretical transmissivity (due to absorption only), i.e., column~2 divided by column~3, is given in column~4. From this excess loss of transmissivity due to surface scattering, $(1-L_\text{t})^n$, the roughness parameter $b$ was calculated.

The roughness parameters of single foils extracted from the UCN transmission measurements of Atchison et al.~\cite{atchison:2009-foils} are consistent between the individual samples and yield average values of $b\text{(Al)}=195\pm 36$~\AA\, for aluminum and of $b\text{(Zr)}=97\pm 19$~\AA{} for zirconium. These are in line with the typical roughness parameters from Steyerl~\cite{steyerl:1972} and prove conclusively that surface scattering is the reason for the measured 2.2-fold and 2.6-fold decrease of foil transmissivity for aluminum and zirconium, respectively, compared with theory (taking into account only absorption), as reported by Atchison et al.~\cite{atchison:2009-foils}.

\section{Conclusion}

In our experiments, we have demonstrated how neutron scattering in the sample bulk can be separated from scattering at the sample surface. We found that ultracold neutrons (UCN) are very susceptible to surface scattering -- an effect that should be taken into account when planning transmission experiments with UCNs. Our results show conclusively that low-roughness sample containers, such as the one presented previously for cryogenic samples~\cite{doege:2018}, must be used to increase the accuracy of results in UCN experiments.

In particular, we found that off-the-shelf high-purity metal foils have the following roughness parameters as seen by UCNs: Al $182\pm 8$~\AA{}, Cu $147\pm 12$~\AA{}, and Zr $313\pm 64$~\AA{}.

Comparing the surface roughness extracted from UCN measurements with that measured by AFM leads to the assumption that neutrons ``see'' a different spectrum of roughness amplitudes than mechanical or optical means of measurement. The results obtained with both methods are, however, of the same order of magnitude.

Applying the method described here, we re-analyzed the UCN transmission data of various metal foils from Atchison et al.~\cite{atchison:2009-foils} and were able to calculate the surface roughness of their sample foils and identify surface scattering as the cause of the 2.2-fold and 2.6-fold decrease of foil transmissivity for aluminum and zirconium, respectively, compared with theory.

The results for this paper were produced as part of the Ph.D. thesis of Stefan D\"oge~\cite{doege:2019-phd}.

\begin{acknowledgments}
We thank Yuliya E. Gorshkova of JINR Dubna for her patient support during the AFM measurements of the metal foils. Support from the reactor crew, the instrument scientists, and technicians of the mechanical workshops at the Institut Laue--Langevin, Grenoble, during the beamtime no. 3-14-380 is gratefully acknowledged. This work received funding from the Russian Foundation for Basic Research (RFBR) under grants no. 17-32-50024-mol-nr and 18-29-19039, from Dr.-Ing. Leonhard-Lorenz-Stiftung, Munich, under grant no. 940/17, and from FRM~II/ Heinz Maier-Leibnitz Zentrum (MLZ), Munich, Germany. The open access publication fee was paid for by the University Library and the Physics Department of the Technische Universit\"at M\"unchen, Munich, Germany.
\end{acknowledgments}

\bibliography{Bibliography}

\begin{thebibliography}{34}%
\makeatletter
\providecommand \@ifxundefined [1]{%
 \@ifx{#1\undefined}
}%
\providecommand \@ifnum [1]{%
 \ifnum #1\expandafter \@firstoftwo
 \else \expandafter \@secondoftwo
 \fi
}%
\providecommand \@ifx [1]{%
 \ifx #1\expandafter \@firstoftwo
 \else \expandafter \@secondoftwo
 \fi
}%
\providecommand \natexlab [1]{#1}%
\providecommand \enquote  [1]{``#1''}%
\providecommand \bibnamefont  [1]{#1}%
\providecommand \bibfnamefont [1]{#1}%
\providecommand \citenamefont [1]{#1}%
\providecommand \href@noop [0]{\@secondoftwo}%
\providecommand \href [0]{\begingroup \@sanitize@url \@href}%
\providecommand \@href[1]{\@@startlink{#1}\@@href}%
\providecommand \@@href[1]{\endgroup#1\@@endlink}%
\providecommand \@sanitize@url [0]{\catcode `\\12\catcode `\$12\catcode
  `\&12\catcode `\#12\catcode `\^12\catcode `\_12\catcode `\%12\relax}%
\providecommand \@@startlink[1]{}%
\providecommand \@@endlink[0]{}%
\providecommand \url  [0]{\begingroup\@sanitize@url \@url }%
\providecommand \@url [1]{\endgroup\@href {#1}{\urlprefix }}%
\providecommand \urlprefix  [0]{URL }%
\providecommand \Eprint [0]{\href }%
\providecommand \doibase [0]{https://doi.org/}%
\providecommand \selectlanguage [0]{\@gobble}%
\providecommand \bibinfo  [0]{\@secondoftwo}%
\providecommand \bibfield  [0]{\@secondoftwo}%
\providecommand \translation [1]{[#1]}%
\providecommand \BibitemOpen [0]{}%
\providecommand \bibitemStop [0]{}%
\providecommand \bibitemNoStop [0]{.\EOS\space}%
\providecommand \EOS [0]{\spacefactor3000\relax}%
\providecommand \BibitemShut  [1]{\csname bibitem#1\endcsname}%
\let\auto@bib@innerbib\@empty
\bibitem [{\citenamefont {Steyerl}\ \emph {et~al.}(1986)\citenamefont
  {Steyerl}, \citenamefont {Nagel}, \citenamefont {Schreiber}, \citenamefont
  {Steinhauser}, \citenamefont {G\"{a}hler}, \citenamefont {Gl\"{a}ser},
  \citenamefont {Ageron}, \citenamefont {Astruc}, \citenamefont {Drexel},
  \citenamefont {Gervais},\ and\ \citenamefont {Mampe}}]{steyerl:1986}%
  \BibitemOpen
  \bibfield  {author} {\bibinfo {author} {\bibfnamefont {A.}~\bibnamefont
  {Steyerl}}, \bibinfo {author} {\bibfnamefont {H.}~\bibnamefont {Nagel}},
  \bibinfo {author} {\bibfnamefont {F.-X.}\ \bibnamefont {Schreiber}}, \bibinfo
  {author} {\bibfnamefont {K.-A.}\ \bibnamefont {Steinhauser}}, \bibinfo
  {author} {\bibfnamefont {R.}~\bibnamefont {G\"{a}hler}}, \bibinfo {author}
  {\bibfnamefont {W.}~\bibnamefont {Gl\"{a}ser}}, \bibinfo {author}
  {\bibfnamefont {P.}~\bibnamefont {Ageron}}, \bibinfo {author} {\bibfnamefont
  {J.~M.}\ \bibnamefont {Astruc}}, \bibinfo {author} {\bibfnamefont
  {W.}~\bibnamefont {Drexel}}, \bibinfo {author} {\bibfnamefont
  {G.}~\bibnamefont {Gervais}},\ and\ \bibinfo {author} {\bibfnamefont
  {W.}~\bibnamefont {Mampe}},\ }\bibfield  {title} {\bibinfo {title} {A new
  source of cold and ultracold neutrons},\ }\href
  {https://doi.org/10.1016/0375-9601(86)90587-6} {\bibfield  {journal}
  {\bibinfo  {journal} {Physics Letters A}\ }\textbf {\bibinfo {volume}
  {116}},\ \bibinfo {pages} {347} (\bibinfo {year} {1986})}\BibitemShut
  {NoStop}%
\bibitem [{\citenamefont {D\"{o}ge}\ \emph {et~al.}(2020)\citenamefont
  {D\"{o}ge}, \citenamefont {Hingerl},\ and\ \citenamefont
  {Morkel}}]{doege:2020-turbine}%
  \BibitemOpen
  \bibfield  {author} {\bibinfo {author} {\bibfnamefont {S.}~\bibnamefont
  {D\"{o}ge}}, \bibinfo {author} {\bibfnamefont {J.}~\bibnamefont {Hingerl}},\
  and\ \bibinfo {author} {\bibfnamefont {C.}~\bibnamefont {Morkel}},\
  }\bibfield  {title} {\bibinfo {title} {Measured velocity spectra and neutron
  densities of the {PF2} ultracold-neutron beam ports at the {I}nstitut
  {L}aue–{L}angevin},\ }\href {https://doi.org/10.1016/j.nima.2019.163112}
  {\bibfield  {journal} {\bibinfo  {journal} {Nuclear Instruments and Methods
  in Physics Research Section A: Accelerators, Spectrometers, Detectors and
  Associated Equipment}\ }\textbf {\bibinfo {volume} {953}},\ \bibinfo {pages}
  {163112} (\bibinfo {year} {2020})}\BibitemShut {NoStop}%
\bibitem [{\citenamefont {Bison}\ \emph {et~al.}(2017)\citenamefont {Bison},
  \citenamefont {Daum}, \citenamefont {Kirch}, \citenamefont {Lauss},
  \citenamefont {Ries}, \citenamefont {Schmidt-Wellenburg}, \citenamefont
  {Zsigmond}, \citenamefont {Brenner}, \citenamefont {Geltenbort},
  \citenamefont {Jenke}, \citenamefont {Zimmer}, \citenamefont {Beck},
  \citenamefont {Heil}, \citenamefont {Kahlenberg}, \citenamefont {Karch},
  \citenamefont {Ross}, \citenamefont {Eberhardt}, \citenamefont {Geppert},
  \citenamefont {Karpuk}, \citenamefont {Reich}, \citenamefont {Siemensen},
  \citenamefont {Sobolev},\ and\ \citenamefont {Trautmann}}]{ries:2017}%
  \BibitemOpen
  \bibfield  {author} {\bibinfo {author} {\bibfnamefont {G.}~\bibnamefont
  {Bison}}, \bibinfo {author} {\bibfnamefont {M.}~\bibnamefont {Daum}},
  \bibinfo {author} {\bibfnamefont {K.}~\bibnamefont {Kirch}}, \bibinfo
  {author} {\bibfnamefont {B.}~\bibnamefont {Lauss}}, \bibinfo {author}
  {\bibfnamefont {D.}~\bibnamefont {Ries}}, \bibinfo {author} {\bibfnamefont
  {P.}~\bibnamefont {Schmidt-Wellenburg}}, \bibinfo {author} {\bibfnamefont
  {G.}~\bibnamefont {Zsigmond}}, \bibinfo {author} {\bibfnamefont
  {T.}~\bibnamefont {Brenner}}, \bibinfo {author} {\bibfnamefont
  {P.}~\bibnamefont {Geltenbort}}, \bibinfo {author} {\bibfnamefont
  {T.}~\bibnamefont {Jenke}}, \bibinfo {author} {\bibfnamefont
  {O.}~\bibnamefont {Zimmer}}, \bibinfo {author} {\bibfnamefont
  {M.}~\bibnamefont {Beck}}, \bibinfo {author} {\bibfnamefont {W.}~\bibnamefont
  {Heil}}, \bibinfo {author} {\bibfnamefont {J.}~\bibnamefont {Kahlenberg}},
  \bibinfo {author} {\bibfnamefont {J.}~\bibnamefont {Karch}}, \bibinfo
  {author} {\bibfnamefont {K.}~\bibnamefont {Ross}}, \bibinfo {author}
  {\bibfnamefont {K.}~\bibnamefont {Eberhardt}}, \bibinfo {author}
  {\bibfnamefont {C.}~\bibnamefont {Geppert}}, \bibinfo {author} {\bibfnamefont
  {S.}~\bibnamefont {Karpuk}}, \bibinfo {author} {\bibfnamefont
  {T.}~\bibnamefont {Reich}}, \bibinfo {author} {\bibfnamefont
  {C.}~\bibnamefont {Siemensen}}, \bibinfo {author} {\bibfnamefont
  {Y.}~\bibnamefont {Sobolev}},\ and\ \bibinfo {author} {\bibfnamefont
  {N.}~\bibnamefont {Trautmann}},\ }\bibfield  {title} {\bibinfo {title}
  {Comparison of ultracold neutron sources for fundamental physics
  measurements},\ }\href {https://doi.org/10.1103/PhysRevC.95.045503}
  {\bibfield  {journal} {\bibinfo  {journal} {Phys. Rev. C}\ }\textbf {\bibinfo
  {volume} {95}},\ \bibinfo {pages} {045503} (\bibinfo {year}
  {2017})}\BibitemShut {NoStop}%
\bibitem [{\citenamefont {Ito}\ \emph {et~al.}(2018)\citenamefont {Ito},
  \citenamefont {Adamek}, \citenamefont {Callahan}, \citenamefont {Choi},
  \citenamefont {Clayton}, \citenamefont {Cude-Woods}, \citenamefont {Currie},
  \citenamefont {Ding}, \citenamefont {Fellers}, \citenamefont {Geltenbort},
  \citenamefont {Lamoreaux}, \citenamefont {Liu}, \citenamefont {MacDonald},
  \citenamefont {Makela}, \citenamefont {Morris}, \citenamefont {Pattie},
  \citenamefont {Ramsey}, \citenamefont {Salvat}, \citenamefont {Saunders},
  \citenamefont {Sharapov}, \citenamefont {Sjue}, \citenamefont {Sprow},
  \citenamefont {Tang}, \citenamefont {Weaver}, \citenamefont {Wei},\ and\
  \citenamefont {Young}}]{ito:2018}%
  \BibitemOpen
  \bibfield  {author} {\bibinfo {author} {\bibfnamefont {T.~M.}\ \bibnamefont
  {Ito}}, \bibinfo {author} {\bibfnamefont {E.~R.}\ \bibnamefont {Adamek}},
  \bibinfo {author} {\bibfnamefont {N.~B.}\ \bibnamefont {Callahan}}, \bibinfo
  {author} {\bibfnamefont {J.~H.}\ \bibnamefont {Choi}}, \bibinfo {author}
  {\bibfnamefont {S.~M.}\ \bibnamefont {Clayton}}, \bibinfo {author}
  {\bibfnamefont {C.}~\bibnamefont {Cude-Woods}}, \bibinfo {author}
  {\bibfnamefont {S.}~\bibnamefont {Currie}}, \bibinfo {author} {\bibfnamefont
  {X.}~\bibnamefont {Ding}}, \bibinfo {author} {\bibfnamefont {D.~E.}\
  \bibnamefont {Fellers}}, \bibinfo {author} {\bibfnamefont {P.}~\bibnamefont
  {Geltenbort}}, \bibinfo {author} {\bibfnamefont {S.~K.}\ \bibnamefont
  {Lamoreaux}}, \bibinfo {author} {\bibfnamefont {C.-Y.}\ \bibnamefont {Liu}},
  \bibinfo {author} {\bibfnamefont {S.}~\bibnamefont {MacDonald}}, \bibinfo
  {author} {\bibfnamefont {M.}~\bibnamefont {Makela}}, \bibinfo {author}
  {\bibfnamefont {C.~L.}\ \bibnamefont {Morris}}, \bibinfo {author}
  {\bibfnamefont {R.~W.}\ \bibnamefont {Pattie}}, \bibinfo {author}
  {\bibfnamefont {J.~C.}\ \bibnamefont {Ramsey}}, \bibinfo {author}
  {\bibfnamefont {D.~J.}\ \bibnamefont {Salvat}}, \bibinfo {author}
  {\bibfnamefont {A.}~\bibnamefont {Saunders}}, \bibinfo {author}
  {\bibfnamefont {E.~I.}\ \bibnamefont {Sharapov}}, \bibinfo {author}
  {\bibfnamefont {S.}~\bibnamefont {Sjue}}, \bibinfo {author} {\bibfnamefont
  {A.~P.}\ \bibnamefont {Sprow}}, \bibinfo {author} {\bibfnamefont
  {Z.}~\bibnamefont {Tang}}, \bibinfo {author} {\bibfnamefont {H.~L.}\
  \bibnamefont {Weaver}}, \bibinfo {author} {\bibfnamefont {W.}~\bibnamefont
  {Wei}},\ and\ \bibinfo {author} {\bibfnamefont {A.~R.}\ \bibnamefont
  {Young}},\ }\bibfield  {title} {\bibinfo {title} {Performance of the upgraded
  ultracold neutron source at {L}os {A}lamos {N}ational {L}aboratory and its
  implication for a possible neutron electric dipole moment experiment},\
  }\href {https://doi.org/10.1103/PhysRevC.97.012501} {\bibfield  {journal}
  {\bibinfo  {journal} {Phys. Rev. C}\ }\textbf {\bibinfo {volume} {97}},\
  \bibinfo {pages} {012501} (\bibinfo {year} {2018})}\BibitemShut {NoStop}%
\bibitem [{\citenamefont {Ahmed}\ \emph {et~al.}(2019)\citenamefont {Ahmed},
  \citenamefont {Altiere}, \citenamefont {Andalib}, \citenamefont {Bell},
  \citenamefont {Bidinosti}, \citenamefont {Cudmore}, \citenamefont {Das},
  \citenamefont {Davis}, \citenamefont {Franke}, \citenamefont {Gericke},
  \citenamefont {Giampa}, \citenamefont {Gnyp}, \citenamefont {Hansen-Romu},
  \citenamefont {Hatanaka}, \citenamefont {Hayamizu}, \citenamefont {Jamieson},
  \citenamefont {Jones}, \citenamefont {Kawasaki}, \citenamefont {Kikawa},
  \citenamefont {Kitaguchi}, \citenamefont {Klassen}, \citenamefont {Konaka},
  \citenamefont {Korkmaz}, \citenamefont {Kuchler}, \citenamefont {Lang},
  \citenamefont {Lee}, \citenamefont {Lindner}, \citenamefont {Madison},
  \citenamefont {Makida}, \citenamefont {Mammei}, \citenamefont {Mammei},
  \citenamefont {Martin}, \citenamefont {Matsumiya}, \citenamefont {Miller},
  \citenamefont {Mishima}, \citenamefont {Momose}, \citenamefont {Okamura},
  \citenamefont {Page}, \citenamefont {Picker}, \citenamefont {Pierre},
  \citenamefont {Ramsay}, \citenamefont {Rebenitsch}, \citenamefont {Rehm},
  \citenamefont {Schreyer}, \citenamefont {Shimizu}, \citenamefont {Sidhu},
  \citenamefont {Sikora}, \citenamefont {Smith}, \citenamefont {Tanihata},
  \citenamefont {Thorsteinson}, \citenamefont {Vanbergen}, \citenamefont {van
  Oers},\ and\ \citenamefont {Watanabe}}]{ahmed:2019}%
  \BibitemOpen
  \bibfield  {author} {\bibinfo {author} {\bibfnamefont {S.}~\bibnamefont
  {Ahmed}}, \bibinfo {author} {\bibfnamefont {E.}~\bibnamefont {Altiere}},
  \bibinfo {author} {\bibfnamefont {T.}~\bibnamefont {Andalib}}, \bibinfo
  {author} {\bibfnamefont {B.}~\bibnamefont {Bell}}, \bibinfo {author}
  {\bibfnamefont {C.~P.}\ \bibnamefont {Bidinosti}}, \bibinfo {author}
  {\bibfnamefont {E.}~\bibnamefont {Cudmore}}, \bibinfo {author} {\bibfnamefont
  {M.}~\bibnamefont {Das}}, \bibinfo {author} {\bibfnamefont {C.~A.}\
  \bibnamefont {Davis}}, \bibinfo {author} {\bibfnamefont {B.}~\bibnamefont
  {Franke}}, \bibinfo {author} {\bibfnamefont {M.}~\bibnamefont {Gericke}},
  \bibinfo {author} {\bibfnamefont {P.}~\bibnamefont {Giampa}}, \bibinfo
  {author} {\bibfnamefont {P.}~\bibnamefont {Gnyp}}, \bibinfo {author}
  {\bibfnamefont {S.}~\bibnamefont {Hansen-Romu}}, \bibinfo {author}
  {\bibfnamefont {K.}~\bibnamefont {Hatanaka}}, \bibinfo {author}
  {\bibfnamefont {T.}~\bibnamefont {Hayamizu}}, \bibinfo {author}
  {\bibfnamefont {B.}~\bibnamefont {Jamieson}}, \bibinfo {author}
  {\bibfnamefont {D.}~\bibnamefont {Jones}}, \bibinfo {author} {\bibfnamefont
  {S.}~\bibnamefont {Kawasaki}}, \bibinfo {author} {\bibfnamefont
  {T.}~\bibnamefont {Kikawa}}, \bibinfo {author} {\bibfnamefont
  {M.}~\bibnamefont {Kitaguchi}}, \bibinfo {author} {\bibfnamefont
  {W.}~\bibnamefont {Klassen}}, \bibinfo {author} {\bibfnamefont
  {A.}~\bibnamefont {Konaka}}, \bibinfo {author} {\bibfnamefont
  {E.}~\bibnamefont {Korkmaz}}, \bibinfo {author} {\bibfnamefont
  {F.}~\bibnamefont {Kuchler}}, \bibinfo {author} {\bibfnamefont
  {M.}~\bibnamefont {Lang}}, \bibinfo {author} {\bibfnamefont {L.}~\bibnamefont
  {Lee}}, \bibinfo {author} {\bibfnamefont {T.}~\bibnamefont {Lindner}},
  \bibinfo {author} {\bibfnamefont {K.~W.}\ \bibnamefont {Madison}}, \bibinfo
  {author} {\bibfnamefont {Y.}~\bibnamefont {Makida}}, \bibinfo {author}
  {\bibfnamefont {J.}~\bibnamefont {Mammei}}, \bibinfo {author} {\bibfnamefont
  {R.}~\bibnamefont {Mammei}}, \bibinfo {author} {\bibfnamefont {J.~W.}\
  \bibnamefont {Martin}}, \bibinfo {author} {\bibfnamefont {R.}~\bibnamefont
  {Matsumiya}}, \bibinfo {author} {\bibfnamefont {E.}~\bibnamefont {Miller}},
  \bibinfo {author} {\bibfnamefont {K.}~\bibnamefont {Mishima}}, \bibinfo
  {author} {\bibfnamefont {T.}~\bibnamefont {Momose}}, \bibinfo {author}
  {\bibfnamefont {T.}~\bibnamefont {Okamura}}, \bibinfo {author} {\bibfnamefont
  {S.}~\bibnamefont {Page}}, \bibinfo {author} {\bibfnamefont {R.}~\bibnamefont
  {Picker}}, \bibinfo {author} {\bibfnamefont {E.}~\bibnamefont {Pierre}},
  \bibinfo {author} {\bibfnamefont {W.~D.}\ \bibnamefont {Ramsay}}, \bibinfo
  {author} {\bibfnamefont {L.}~\bibnamefont {Rebenitsch}}, \bibinfo {author}
  {\bibfnamefont {F.}~\bibnamefont {Rehm}}, \bibinfo {author} {\bibfnamefont
  {W.}~\bibnamefont {Schreyer}}, \bibinfo {author} {\bibfnamefont {H.~M.}\
  \bibnamefont {Shimizu}}, \bibinfo {author} {\bibfnamefont {S.}~\bibnamefont
  {Sidhu}}, \bibinfo {author} {\bibfnamefont {A.}~\bibnamefont {Sikora}},
  \bibinfo {author} {\bibfnamefont {J.}~\bibnamefont {Smith}}, \bibinfo
  {author} {\bibfnamefont {I.}~\bibnamefont {Tanihata}}, \bibinfo {author}
  {\bibfnamefont {B.}~\bibnamefont {Thorsteinson}}, \bibinfo {author}
  {\bibfnamefont {S.}~\bibnamefont {Vanbergen}}, \bibinfo {author}
  {\bibfnamefont {W.~T.~H.}\ \bibnamefont {van Oers}},\ and\ \bibinfo {author}
  {\bibfnamefont {Y.~X.}\ \bibnamefont {Watanabe}} (\bibinfo {collaboration}
  {TUCAN Collaboration}),\ }\bibfield  {title} {\bibinfo {title} {First
  ultracold neutrons produced at {TRIUMF}},\ }\href
  {https://doi.org/10.1103/PhysRevC.99.025503} {\bibfield  {journal} {\bibinfo
  {journal} {Phys. Rev. C}\ }\textbf {\bibinfo {volume} {99}},\ \bibinfo
  {pages} {025503} (\bibinfo {year} {2019})}\BibitemShut {NoStop}%
\bibitem [{\citenamefont {Steyerl}(1972)}]{steyerl:1972}%
  \BibitemOpen
  \bibfield  {author} {\bibinfo {author} {\bibfnamefont {A.}~\bibnamefont
  {Steyerl}},\ }\bibfield  {title} {\bibinfo {title} {Effect of surface
  roughness on the total reflexion and transmission of slow neutrons},\ }\href
  {https://doi.org/10.1007/BF01380066} {\bibfield  {journal} {\bibinfo
  {journal} {Zeitschrift f{\"u}r Physik A Hadrons and Nuclei}\ }\textbf
  {\bibinfo {volume} {254}},\ \bibinfo {pages} {169} (\bibinfo {year}
  {1972})}\BibitemShut {NoStop}%
\bibitem [{\citenamefont {Roth}(1977)}]{roth:1977}%
  \BibitemOpen
  \bibfield  {author} {\bibinfo {author} {\bibfnamefont {M.}~\bibnamefont
  {Roth}},\ }\bibfield  {title} {\bibinfo {title} {{The small-angle scattering
  of neutrons by surface imperfections}},\ }\href
  {https://doi.org/10.1107/S0021889877013181} {\bibfield  {journal} {\bibinfo
  {journal} {Journal of Applied Crystallography}\ }\textbf {\bibinfo {volume}
  {10}},\ \bibinfo {pages} {172} (\bibinfo {year} {1977})}\BibitemShut
  {NoStop}%
\bibitem [{\citenamefont {Frei}(2002)}]{frei:2002}%
  \BibitemOpen
  \bibfield  {author} {\bibinfo {author} {\bibfnamefont {A.}~\bibnamefont
  {Frei}},\ }\emph {\bibinfo {title} {{U}ntersuchung ausgew\"{a}hlter
  {E}lemente der {UCN}-{Q}uelle {M}ini-{D}2}},\ \href@noop {} {\bibinfo {type}
  {Diploma thesis}},\ \bibinfo  {school} {Technische Universit\"at M\"unchen,
  Munich, Germany} (\bibinfo {year} {2002})\BibitemShut {NoStop}%
\bibitem [{\citenamefont {Atchison}\ \emph {et~al.}(2009)\citenamefont
  {Atchison}, \citenamefont {Blau}, \citenamefont {Bollhalder}, \citenamefont
  {Daum}, \citenamefont {Fierlinger}, \citenamefont {Geltenbort}, \citenamefont
  {Hampel}, \citenamefont {Kasprzak}, \citenamefont {Kirch}, \citenamefont
  {K{\"{o}}chli}, \citenamefont {Kuczewski}, \citenamefont {Leber},
  \citenamefont {Locher}, \citenamefont {Meier}, \citenamefont {Ochse},
  \citenamefont {Pichlmaier}, \citenamefont {Plonka}, \citenamefont {Reiser},
  \citenamefont {Ulrich}, \citenamefont {Wang}, \citenamefont {Wiehl},
  \citenamefont {Zimmer},\ and\ \citenamefont
  {Zsigmond}}]{atchison:2009-foils}%
  \BibitemOpen
  \bibfield  {author} {\bibinfo {author} {\bibfnamefont {F.}~\bibnamefont
  {Atchison}}, \bibinfo {author} {\bibfnamefont {B.}~\bibnamefont {Blau}},
  \bibinfo {author} {\bibfnamefont {A.}~\bibnamefont {Bollhalder}}, \bibinfo
  {author} {\bibfnamefont {M.}~\bibnamefont {Daum}}, \bibinfo {author}
  {\bibfnamefont {P.}~\bibnamefont {Fierlinger}}, \bibinfo {author}
  {\bibfnamefont {P.}~\bibnamefont {Geltenbort}}, \bibinfo {author}
  {\bibfnamefont {G.}~\bibnamefont {Hampel}}, \bibinfo {author} {\bibfnamefont
  {M.}~\bibnamefont {Kasprzak}}, \bibinfo {author} {\bibfnamefont
  {K.}~\bibnamefont {Kirch}}, \bibinfo {author} {\bibfnamefont
  {S.}~\bibnamefont {K{\"{o}}chli}}, \bibinfo {author} {\bibfnamefont
  {B.}~\bibnamefont {Kuczewski}}, \bibinfo {author} {\bibfnamefont
  {H.}~\bibnamefont {Leber}}, \bibinfo {author} {\bibfnamefont
  {M.}~\bibnamefont {Locher}}, \bibinfo {author} {\bibfnamefont
  {M.}~\bibnamefont {Meier}}, \bibinfo {author} {\bibfnamefont
  {S.}~\bibnamefont {Ochse}}, \bibinfo {author} {\bibfnamefont
  {A.}~\bibnamefont {Pichlmaier}}, \bibinfo {author} {\bibfnamefont
  {C.}~\bibnamefont {Plonka}}, \bibinfo {author} {\bibfnamefont
  {R.}~\bibnamefont {Reiser}}, \bibinfo {author} {\bibfnamefont
  {J.}~\bibnamefont {Ulrich}}, \bibinfo {author} {\bibfnamefont
  {X.}~\bibnamefont {Wang}}, \bibinfo {author} {\bibfnamefont {N.}~\bibnamefont
  {Wiehl}}, \bibinfo {author} {\bibfnamefont {O.}~\bibnamefont {Zimmer}},\ and\
  \bibinfo {author} {\bibfnamefont {G.}~\bibnamefont {Zsigmond}},\ }\bibfield
  {title} {\bibinfo {title} {Transmission of very slow neutrons through
  material foils and its influence on the design of ultracold neutron
  sources},\ }\href {https://doi.org/10.1016/j.nima.2009.06.047} {\bibfield
  {journal} {\bibinfo  {journal} {Nuclear Instruments and Methods in Physics
  Research Section A: Accelerators, Spectrometers, Detectors and Associated
  Equipment}\ }\textbf {\bibinfo {volume} {608}},\ \bibinfo {pages} {144}
  (\bibinfo {year} {2009})}\BibitemShut {NoStop}%
\bibitem [{\citenamefont {Lavelle}\ \emph {et~al.}(2010)\citenamefont
  {Lavelle}, \citenamefont {Liu}, \citenamefont {Fox}, \citenamefont {Manus},
  \citenamefont {McChesney}, \citenamefont {Salvat}, \citenamefont {Shin},
  \citenamefont {Makela}, \citenamefont {Morris}, \citenamefont {Saunders},
  \citenamefont {Couture},\ and\ \citenamefont {Young}}]{lavelle:2010}%
  \BibitemOpen
  \bibfield  {author} {\bibinfo {author} {\bibfnamefont {C.~M.}\ \bibnamefont
  {Lavelle}}, \bibinfo {author} {\bibfnamefont {C.-Y.}\ \bibnamefont {Liu}},
  \bibinfo {author} {\bibfnamefont {W.}~\bibnamefont {Fox}}, \bibinfo {author}
  {\bibfnamefont {G.}~\bibnamefont {Manus}}, \bibinfo {author} {\bibfnamefont
  {P.~M.}\ \bibnamefont {McChesney}}, \bibinfo {author} {\bibfnamefont {D.~J.}\
  \bibnamefont {Salvat}}, \bibinfo {author} {\bibfnamefont {Y.}~\bibnamefont
  {Shin}}, \bibinfo {author} {\bibfnamefont {M.}~\bibnamefont {Makela}},
  \bibinfo {author} {\bibfnamefont {C.}~\bibnamefont {Morris}}, \bibinfo
  {author} {\bibfnamefont {A.}~\bibnamefont {Saunders}}, \bibinfo {author}
  {\bibfnamefont {A.}~\bibnamefont {Couture}},\ and\ \bibinfo {author}
  {\bibfnamefont {A.~R.}\ \bibnamefont {Young}},\ }\bibfield  {title} {\bibinfo
  {title} {Ultracold-neutron production in a pulsed-neutron beam line},\ }\href
  {https://doi.org/10.1103/PhysRevC.82.015502} {\bibfield  {journal} {\bibinfo
  {journal} {Phys. Rev. C}\ }\textbf {\bibinfo {volume} {82}},\ \bibinfo
  {pages} {015502} (\bibinfo {year} {2010})}\BibitemShut {NoStop}%
\bibitem [{\citenamefont {D\"{o}ge}\ and\ \citenamefont
  {Hingerl}(2018)}]{doege:2018}%
  \BibitemOpen
  \bibfield  {author} {\bibinfo {author} {\bibfnamefont {S.}~\bibnamefont
  {D\"{o}ge}}\ and\ \bibinfo {author} {\bibfnamefont {J.}~\bibnamefont
  {Hingerl}},\ }\bibfield  {title} {\bibinfo {title} {A hydrogen leak-tight,
  transparent cryogenic sample container for ultracold-neutron transmission
  measurements},\ }\href {https://doi.org/10.1063/1.4996296} {\bibfield
  {journal} {\bibinfo  {journal} {Review of Scientific Instruments}\ }\textbf
  {\bibinfo {volume} {89}},\ \bibinfo {pages} {033903} (\bibinfo {year}
  {2018})},\ \Eprint {https://arxiv.org/abs/1803.10159} {1803.10159}
  \BibitemShut {NoStop}%
\bibitem [{\citenamefont {Ignatovich}(1990)}]{ignatovich:1990}%
  \BibitemOpen
  \bibfield  {author} {\bibinfo {author} {\bibfnamefont {V.~K.}\ \bibnamefont
  {Ignatovich}},\ }\href@noop {} {\emph {\bibinfo {title} {The Physics of
  Ultracold Neutrons}}}\ (\bibinfo  {publisher} {Clarendon Press, Oxford},\
  \bibinfo {year} {1990})\BibitemShut {NoStop}%
\bibitem [{\citenamefont {Pokotilovski}(1999)}]{pokotilovski:1999}%
  \BibitemOpen
  \bibfield  {author} {\bibinfo {author} {\bibfnamefont {Y.~N.}\ \bibnamefont
  {Pokotilovski}},\ }\bibfield  {title} {\bibinfo {title} {Interaction of
  ultracold neutrons with liquid surface modes as a possible reason for neutron
  energy spread during long storage in fluid wall traps},\ }\href
  {https://doi.org/10.1016/S0375-9601(99)00122-X} {\bibfield  {journal}
  {\bibinfo  {journal} {Physics Letters A}\ }\textbf {\bibinfo {volume}
  {255}},\ \bibinfo {pages} {173} (\bibinfo {year} {1999})}\BibitemShut
  {NoStop}%
\bibitem [{\citenamefont {Strelkov}\ and\ \citenamefont
  {Hetzelt}(1978)}]{strelkov:1978}%
  \BibitemOpen
  \bibfield  {author} {\bibinfo {author} {\bibfnamefont {A.~V.}\ \bibnamefont
  {Strelkov}}\ and\ \bibinfo {author} {\bibfnamefont {M.}~\bibnamefont
  {Hetzelt}},\ }\bibfield  {title} {\bibinfo {title} {Observation of heating of
  ultracold neutrons as the cause of the anomalous limitation of their
  confinement time in closed vessels},\ }\href
  {http://www.jetp.ac.ru/cgi-bin/dn/e_047_01_0011.pdf} {\bibfield  {journal}
  {\bibinfo  {journal} {Sov. Phys. JETP}\ }\textbf {\bibinfo {volume} {47}},\
  \bibinfo {pages} {11} (\bibinfo {year} {1978})}\BibitemShut {NoStop}%
\bibitem [{\citenamefont {Lychagin}\ \emph {et~al.}(2000)\citenamefont
  {Lychagin}, \citenamefont {Muzychka}, \citenamefont {Nesvizhevsky},
  \citenamefont {Nekhaev}, \citenamefont {Tal’daev},\ and\ \citenamefont
  {Strelkov}}]{lychagin:2000}%
  \BibitemOpen
  \bibfield  {author} {\bibinfo {author} {\bibfnamefont {E.~V.}\ \bibnamefont
  {Lychagin}}, \bibinfo {author} {\bibfnamefont {A.~Y.}\ \bibnamefont
  {Muzychka}}, \bibinfo {author} {\bibfnamefont {V.~V.}\ \bibnamefont
  {Nesvizhevsky}}, \bibinfo {author} {\bibfnamefont {G.~V.}\ \bibnamefont
  {Nekhaev}}, \bibinfo {author} {\bibfnamefont {R.~R.}\ \bibnamefont
  {Tal’daev}},\ and\ \bibinfo {author} {\bibfnamefont {A.~V.}\ \bibnamefont
  {Strelkov}},\ }\bibfield  {title} {\bibinfo {title} {Temperature dependence
  of inelastic ultracold-neutron scattering at low energy transfer},\ }\href
  {https://doi.org/10.1134/1.855666} {\bibfield  {journal} {\bibinfo  {journal}
  {Physics of Atomic Nuclei}\ }\textbf {\bibinfo {volume} {63}},\ \bibinfo
  {pages} {548} (\bibinfo {year} {2000})}\BibitemShut {NoStop}%
\bibitem [{\citenamefont {Steyerl}\ \emph {et~al.}(2010)\citenamefont
  {Steyerl}, \citenamefont {Malik}, \citenamefont {Desai},\ and\ \citenamefont
  {Kaufman}}]{steyerl:2010}%
  \BibitemOpen
  \bibfield  {author} {\bibinfo {author} {\bibfnamefont {A.}~\bibnamefont
  {Steyerl}}, \bibinfo {author} {\bibfnamefont {S.~S.}\ \bibnamefont {Malik}},
  \bibinfo {author} {\bibfnamefont {A.~M.}\ \bibnamefont {Desai}},\ and\
  \bibinfo {author} {\bibfnamefont {C.}~\bibnamefont {Kaufman}},\ }\bibfield
  {title} {\bibinfo {title} {Surface roughness effect on ultracold neutron
  interaction with a wall and implications for computer simulations},\ }\href
  {https://doi.org/10.1103/PhysRevC.81.055505} {\bibfield  {journal} {\bibinfo
  {journal} {Phys. Rev. C}\ }\textbf {\bibinfo {volume} {81}},\ \bibinfo
  {pages} {055505} (\bibinfo {year} {2010})}\BibitemShut {NoStop}%
\bibitem [{\citenamefont {Sears}(1992)}]{sears:1992}%
  \BibitemOpen
  \bibfield  {author} {\bibinfo {author} {\bibfnamefont {V.~F.}\ \bibnamefont
  {Sears}},\ }\bibfield  {title} {\bibinfo {title} {Neutron scattering lengths
  and cross sections},\ }\href {http://www.ncnr.nist.gov/resources/n-lengths/}
  {\bibfield  {journal} {\bibinfo  {journal} {Neutron News}\ }\textbf {\bibinfo
  {volume} {3}},\ \bibinfo {pages} {26} (\bibinfo {year} {1992})}\BibitemShut
  {NoStop}%
\bibitem [{\citenamefont {Keil}\ \emph {et~al.}(2007)\citenamefont {Keil},
  \citenamefont {{L\"{u}tzenkirchen-Hecht}},\ and\ \citenamefont
  {Frahm}}]{keil:2007}%
  \BibitemOpen
  \bibfield  {author} {\bibinfo {author} {\bibfnamefont {P.}~\bibnamefont
  {Keil}}, \bibinfo {author} {\bibfnamefont {D.}~\bibnamefont
  {{L\"{u}tzenkirchen-Hecht}}},\ and\ \bibinfo {author} {\bibfnamefont
  {R.}~\bibnamefont {Frahm}},\ }\bibfield  {title} {\bibinfo {title}
  {{I}nvestigation of {R}oom {T}emperature {O}xidation of {Cu} in {A}ir by
  {Y}oneda-{XAFS}},\ }\href {https://doi.org/10.1063/1.2644569} {\bibfield
  {journal} {\bibinfo  {journal} {AIP Conference Proceedings}\ }\textbf
  {\bibinfo {volume} {882}},\ \bibinfo {pages} {490} (\bibinfo {year}
  {2007})}\BibitemShut {NoStop}%
\bibitem [{\citenamefont {Platzman}\ \emph {et~al.}(2008)\citenamefont
  {Platzman}, \citenamefont {Brener}, \citenamefont {Haick},\ and\
  \citenamefont {Tannenbaum}}]{platzman:2008}%
  \BibitemOpen
  \bibfield  {author} {\bibinfo {author} {\bibfnamefont {I.}~\bibnamefont
  {Platzman}}, \bibinfo {author} {\bibfnamefont {R.}~\bibnamefont {Brener}},
  \bibinfo {author} {\bibfnamefont {H.}~\bibnamefont {Haick}},\ and\ \bibinfo
  {author} {\bibfnamefont {R.}~\bibnamefont {Tannenbaum}},\ }\bibfield  {title}
  {\bibinfo {title} {{O}xidation of {P}olycrystalline {C}opper {T}hin {F}ilms
  at {A}mbient {C}onditions},\ }\href {https://doi.org/10.1021/jp076981k}
  {\bibfield  {journal} {\bibinfo  {journal} {The Journal of Physical Chemistry
  C}\ }\textbf {\bibinfo {volume} {112}},\ \bibinfo {pages} {1101} (\bibinfo
  {year} {2008})}\BibitemShut {NoStop}%
\bibitem [{\citenamefont {Tamura}\ \emph {et~al.}(2003)\citenamefont {Tamura},
  \citenamefont {Kimura}, \citenamefont {Suzuki}, \citenamefont {Kido},
  \citenamefont {Sato}, \citenamefont {Tanigaki}, \citenamefont {Kurumada},
  \citenamefont {Saito},\ and\ \citenamefont {Kaito}}]{tamura:2003}%
  \BibitemOpen
  \bibfield  {author} {\bibinfo {author} {\bibfnamefont {K.}~\bibnamefont
  {Tamura}}, \bibinfo {author} {\bibfnamefont {Y.}~\bibnamefont {Kimura}},
  \bibinfo {author} {\bibfnamefont {H.}~\bibnamefont {Suzuki}}, \bibinfo
  {author} {\bibfnamefont {O.}~\bibnamefont {Kido}}, \bibinfo {author}
  {\bibfnamefont {T.}~\bibnamefont {Sato}}, \bibinfo {author} {\bibfnamefont
  {T.}~\bibnamefont {Tanigaki}}, \bibinfo {author} {\bibfnamefont
  {M.}~\bibnamefont {Kurumada}}, \bibinfo {author} {\bibfnamefont
  {Y.}~\bibnamefont {Saito}},\ and\ \bibinfo {author} {\bibfnamefont
  {C.}~\bibnamefont {Kaito}},\ }\bibfield  {title} {\bibinfo {title} {Structure
  and thickness of natural oxide layer on ultrafine particle},\ }\href
  {https://doi.org/10.1143/JJAP.42.7489} {\bibfield  {journal} {\bibinfo
  {journal} {Japanese Journal of Applied Physics}\ }\textbf {\bibinfo {volume}
  {42}},\ \bibinfo {pages} {7489} (\bibinfo {year} {2003})}\BibitemShut
  {NoStop}%
\bibitem [{\citenamefont {Evertsson}\ \emph {et~al.}(2015)\citenamefont
  {Evertsson}, \citenamefont {Bertram}, \citenamefont {Zhang}, \citenamefont
  {Rullik}, \citenamefont {Merte}, \citenamefont {Shipilin}, \citenamefont
  {Soldemo}, \citenamefont {Ahmadi}, \citenamefont {Vinogradov}, \citenamefont
  {Carl\`{a}}, \citenamefont {Weissenrieder}, \citenamefont {G{\"{o}}thelid},
  \citenamefont {Pan}, \citenamefont {Mikkelsen}, \citenamefont {Nilsson},\
  and\ \citenamefont {Lundgren}}]{evertsson:2015}%
  \BibitemOpen
  \bibfield  {author} {\bibinfo {author} {\bibfnamefont {J.}~\bibnamefont
  {Evertsson}}, \bibinfo {author} {\bibfnamefont {F.}~\bibnamefont {Bertram}},
  \bibinfo {author} {\bibfnamefont {F.}~\bibnamefont {Zhang}}, \bibinfo
  {author} {\bibfnamefont {L.}~\bibnamefont {Rullik}}, \bibinfo {author}
  {\bibfnamefont {L.~R.}\ \bibnamefont {Merte}}, \bibinfo {author}
  {\bibfnamefont {M.}~\bibnamefont {Shipilin}}, \bibinfo {author}
  {\bibfnamefont {M.}~\bibnamefont {Soldemo}}, \bibinfo {author} {\bibfnamefont
  {S.}~\bibnamefont {Ahmadi}}, \bibinfo {author} {\bibfnamefont
  {N.}~\bibnamefont {Vinogradov}}, \bibinfo {author} {\bibfnamefont
  {F.}~\bibnamefont {Carl\`{a}}}, \bibinfo {author} {\bibfnamefont
  {J.}~\bibnamefont {Weissenrieder}}, \bibinfo {author} {\bibfnamefont
  {M.}~\bibnamefont {G{\"{o}}thelid}}, \bibinfo {author} {\bibfnamefont
  {J.}~\bibnamefont {Pan}}, \bibinfo {author} {\bibfnamefont {A.}~\bibnamefont
  {Mikkelsen}}, \bibinfo {author} {\bibfnamefont {J.-O.}\ \bibnamefont
  {Nilsson}},\ and\ \bibinfo {author} {\bibfnamefont {E.}~\bibnamefont
  {Lundgren}},\ }\bibfield  {title} {\bibinfo {title} {The thickness of native
  oxides on aluminum alloys and single crystals},\ }\href
  {https://doi.org/10.1016/j.apsusc.2015.05.043} {\bibfield  {journal}
  {\bibinfo  {journal} {Applied Surface Science}\ }\textbf {\bibinfo {volume}
  {349}},\ \bibinfo {pages} {826} (\bibinfo {year} {2015})}\BibitemShut
  {NoStop}%
\bibitem [{\citenamefont {Bespalov}\ \emph {et~al.}(2015)\citenamefont
  {Bespalov}, \citenamefont {Datler}, \citenamefont {Buhr}, \citenamefont
  {Drachsel}, \citenamefont {Rupprechter},\ and\ \citenamefont
  {Suchorski}}]{bespalov:2015}%
  \BibitemOpen
  \bibfield  {author} {\bibinfo {author} {\bibfnamefont {I.}~\bibnamefont
  {Bespalov}}, \bibinfo {author} {\bibfnamefont {M.}~\bibnamefont {Datler}},
  \bibinfo {author} {\bibfnamefont {S.}~\bibnamefont {Buhr}}, \bibinfo {author}
  {\bibfnamefont {W.}~\bibnamefont {Drachsel}}, \bibinfo {author}
  {\bibfnamefont {G.}~\bibnamefont {Rupprechter}},\ and\ \bibinfo {author}
  {\bibfnamefont {Y.}~\bibnamefont {Suchorski}},\ }\bibfield  {title} {\bibinfo
  {title} {{I}nitial stages of oxide formation on the {Zr} surface at low
  oxygen pressure: {A}n in situ {FIM} and {XPS} study},\ }\href
  {https://doi.org/10.1016/j.ultramic.2015.02.016} {\bibfield  {journal}
  {\bibinfo  {journal} {Ultramicroscopy}\ }\textbf {\bibinfo {volume} {159}},\
  \bibinfo {pages} {147} (\bibinfo {year} {2015})}\BibitemShut {NoStop}%
\bibitem [{\citenamefont {Pokotilovski}(2016)}]{pokotilovski:2016}%
  \BibitemOpen
  \bibfield  {author} {\bibinfo {author} {\bibfnamefont {Y.~N.}\ \bibnamefont
  {Pokotilovski}},\ }\bibfield  {title} {\bibinfo {title} {{E}ffect of oxide
  films and structural inhomogeneities on transmission of ultracold neutrons
  through foils},\ }\href {https://doi.org/10.1051/epjap/2016150073} {\bibfield
   {journal} {\bibinfo  {journal} {Eur. Phys. J. Appl. Phys.}\ }\textbf
  {\bibinfo {volume} {73}},\ \bibinfo {pages} {20302} (\bibinfo {year}
  {2016})}\BibitemShut {NoStop}%
\bibitem [{\citenamefont {Turchin}(1965)}]{turchin:1965}%
  \BibitemOpen
  \bibfield  {author} {\bibinfo {author} {\bibfnamefont {V.~F.}\ \bibnamefont
  {Turchin}},\ }\href@noop {} {\emph {\bibinfo {title} {Slow Neutrons}}},\
  {I}srael program for scientific translations\ (\bibinfo  {publisher} {Sivan
  Press, Jerusalem},\ \bibinfo {year} {1965})\ \bibinfo {note} {{R}ussian
  original: Medlennye nejtrony (Gosatomizdat, Moscow, 1963)}\BibitemShut
  {NoStop}%
\bibitem [{\citenamefont {Placzek}\ and\ \citenamefont
  {Van~Hove}(1955)}]{placzek:1955}%
  \BibitemOpen
  \bibfield  {author} {\bibinfo {author} {\bibfnamefont {G.}~\bibnamefont
  {Placzek}}\ and\ \bibinfo {author} {\bibfnamefont {L.}~\bibnamefont
  {Van~Hove}},\ }\bibfield  {title} {\bibinfo {title} {Interference effects in
  the total neutron scattering cross-section of crystals},\ }\href
  {https://doi.org/10.1007/BF02731767} {\bibfield  {journal} {\bibinfo
  {journal} {Il Nuovo Cimento (1955-1965)}\ }\textbf {\bibinfo {volume} {1}},\
  \bibinfo {pages} {233} (\bibinfo {year} {1955})}\BibitemShut {NoStop}%
\bibitem [{\citenamefont {Bouguer}(1729)}]{bouguer:1729}%
  \BibitemOpen
  \bibfield  {author} {\bibinfo {author} {\bibfnamefont {P.}~\bibnamefont
  {Bouguer}},\ }\href@noop {} {\emph {\bibinfo {title} {Essai d'optique sur la
  gradation de la lumi\`ere}}}\ (\bibinfo  {publisher} {Claude Jombert,
  Paris},\ \bibinfo {year} {1729})\BibitemShut {NoStop}%
\bibitem [{\citenamefont {Beer}(1852)}]{beer:1852}%
  \BibitemOpen
  \bibfield  {author} {\bibinfo {author} {\bibfnamefont {A.}~\bibnamefont
  {Beer}},\ }\bibfield  {title} {\bibinfo {title} {{B}estimmung der
  {A}bsorption des rothen {L}ichts in farbigen {F}l\"{u}ssigkeiten},\
  }\href@noop {} {\bibfield  {journal} {\bibinfo  {journal} {Annalen der Physik
  und Chemie}\ }\textbf {\bibinfo {volume} {86}},\ \bibinfo {pages} {78}
  (\bibinfo {year} {1852})}\BibitemShut {NoStop}%
\bibitem [{\citenamefont {D\"oge}\ \emph {et~al.}(2015)\citenamefont {D\"oge},
  \citenamefont {Herold}, \citenamefont {M\"uller}, \citenamefont {Morkel},
  \citenamefont {Gutsmiedl}, \citenamefont {Geltenbort}, \citenamefont {Lauer},
  \citenamefont {Fierlinger}, \citenamefont {Petry},\ and\ \citenamefont
  {B\"oni}}]{doege:2015}%
  \BibitemOpen
  \bibfield  {author} {\bibinfo {author} {\bibfnamefont {S.}~\bibnamefont
  {D\"oge}}, \bibinfo {author} {\bibfnamefont {C.}~\bibnamefont {Herold}},
  \bibinfo {author} {\bibfnamefont {S.}~\bibnamefont {M\"uller}}, \bibinfo
  {author} {\bibfnamefont {C.}~\bibnamefont {Morkel}}, \bibinfo {author}
  {\bibfnamefont {E.}~\bibnamefont {Gutsmiedl}}, \bibinfo {author}
  {\bibfnamefont {P.}~\bibnamefont {Geltenbort}}, \bibinfo {author}
  {\bibfnamefont {T.}~\bibnamefont {Lauer}}, \bibinfo {author} {\bibfnamefont
  {P.}~\bibnamefont {Fierlinger}}, \bibinfo {author} {\bibfnamefont
  {W.}~\bibnamefont {Petry}},\ and\ \bibinfo {author} {\bibfnamefont
  {P.}~\bibnamefont {B\"oni}},\ }\bibfield  {title} {\bibinfo {title}
  {Scattering cross sections of liquid deuterium for ultracold neutrons:
  Experimental results and a calculation model},\ }\href
  {https://doi.org/10.1103/PhysRevB.91.214309} {\bibfield  {journal} {\bibinfo
  {journal} {Phys. Rev. B}\ }\textbf {\bibinfo {volume} {91}},\ \bibinfo
  {pages} {214309} (\bibinfo {year} {2015})},\ \Eprint
  {https://arxiv.org/abs/1511.07065} {1511.07065} \BibitemShut {NoStop}%
\bibitem [{\citenamefont {Whitehouse}(2004)}]{whitehouse:2004}%
  \BibitemOpen
  \bibfield  {author} {\bibinfo {author} {\bibfnamefont {D.}~\bibnamefont
  {Whitehouse}},\ }\href@noop {} {\emph {\bibinfo {title} {{S}urfaces and their
  {M}easurement}}}\ (\bibinfo  {publisher} {Butterworth-Heinemann, Boston},\
  \bibinfo {year} {2004})\BibitemShut {NoStop}%
\bibitem [{\citenamefont {Doege}\ \emph {et~al.}(2018)\citenamefont {Doege},
  \citenamefont {Morkel},\ and\ \citenamefont {Hingerl}}]{doege:2018-3-14-380}%
  \BibitemOpen
  \bibfield  {author} {\bibinfo {author} {\bibfnamefont {S.}~\bibnamefont
  {Doege}}, \bibinfo {author} {\bibfnamefont {C.}~\bibnamefont {Morkel}},\ and\
  \bibinfo {author} {\bibfnamefont {J.}~\bibnamefont {Hingerl}},\ }\bibfield
  {title} {\bibinfo {title} {{UCN} scattering on metal surfaces}\ }\href
  {https://doi.org/10.5291/ill-data.3-14-380} {10.5291/ill-data.3-14-380}
  (\bibinfo {year} {2018}),\ \Eprint
  {https://arxiv.org/abs/https://doi.org/10.5291/ill-data.3-14-380}
  {https://doi.org/10.5291/ill-data.3-14-380} \BibitemShut {NoStop}%
\bibitem [{\citenamefont {Poon}\ and\ \citenamefont
  {Bhushan}(1995)}]{poon:1995}%
  \BibitemOpen
  \bibfield  {author} {\bibinfo {author} {\bibfnamefont {C.~Y.}\ \bibnamefont
  {Poon}}\ and\ \bibinfo {author} {\bibfnamefont {B.}~\bibnamefont {Bhushan}},\
  }\bibfield  {title} {\bibinfo {title} {Comparison of surface roughness
  measurements by stylus profiler, {AFM} and non-contact optical profiler},\
  }\href {https://doi.org/10.1016/0043-1648(95)06697-7} {\bibfield  {journal}
  {\bibinfo  {journal} {Wear}\ }\textbf {\bibinfo {volume} {190}},\ \bibinfo
  {pages} {76} (\bibinfo {year} {1995})},\ \bibinfo {note} {{M}acro and
  Micro-Tribology and Mechanics of Magnetic Storage Systems}\BibitemShut
  {NoStop}%
\bibitem [{\citenamefont {Gadelmawla}\ \emph {et~al.}(2002)\citenamefont
  {Gadelmawla}, \citenamefont {Koura}, \citenamefont {Maksoud}, \citenamefont
  {Elewa},\ and\ \citenamefont {Soliman}}]{gadelmawla:2002}%
  \BibitemOpen
  \bibfield  {author} {\bibinfo {author} {\bibfnamefont {E.~S.}\ \bibnamefont
  {Gadelmawla}}, \bibinfo {author} {\bibfnamefont {M.~M.}\ \bibnamefont
  {Koura}}, \bibinfo {author} {\bibfnamefont {T.~M.~A.}\ \bibnamefont
  {Maksoud}}, \bibinfo {author} {\bibfnamefont {I.~M.}\ \bibnamefont {Elewa}},\
  and\ \bibinfo {author} {\bibfnamefont {H.~H.}\ \bibnamefont {Soliman}},\
  }\bibfield  {title} {\bibinfo {title} {Roughness parameters},\ }\href
  {https://doi.org/10.1016/S0924-0136(02)00060-2} {\bibfield  {journal}
  {\bibinfo  {journal} {Journal of Materials Processing Technology}\ }\textbf
  {\bibinfo {volume} {123}},\ \bibinfo {pages} {133} (\bibinfo {year}
  {2002})}\BibitemShut {NoStop}%
\bibitem [{\citenamefont {Daum}(2018)}]{daum:2018}%
  \BibitemOpen
  \bibfield  {author} {\bibinfo {author} {\bibfnamefont {M.}~\bibnamefont
  {Daum}},\ }\bibfield  {title} {\bibinfo {title} {{P}aul {S}cherrer
  {I}nstitut, {V}illigen, {S}witzerland},\ }\href@noop {} {\  (\bibinfo {year}
  {2018})},\ \bibinfo {note} {priv. comm.}\BibitemShut {Stop}%
\bibitem [{\citenamefont {D\"{o}ge}(2019)}]{doege:2019-phd}%
  \BibitemOpen
  \bibfield  {author} {\bibinfo {author} {\bibfnamefont {S.}~\bibnamefont
  {D\"{o}ge}},\ }\emph {\bibinfo {title} {{S}cattering of {U}ltracold
  {N}eutrons in {C}ondensed {D}euterium and on {M}aterial {S}urfaces}},\ \href
  {https://doi.org/10.14459/2019md1464401} {\bibinfo {type} {{PhD} thesis}},\
  \bibinfo  {school} {Technische {U}niversit\"{a}t M\"{u}nchen, Munich,
  Germany} (\bibinfo {year} {2019}),\ \Eprint
  {https://arxiv.org/abs/http://doi.org/10.14459/2019md1464401}
  {http://doi.org/10.14459/2019md1464401} \BibitemShut {NoStop}%
\end{thebibliography}%

\end{document}